\documentclass[prd,nofootinbib,preprint,onecolumn]{revtex4}

\usepackage{amssymb} 
\usepackage{amsmath}
\usepackage{xfrac}
\usepackage{bm}
\usepackage{slashed}
\usepackage{datetime}
\usepackage{braket}
\usepackage[noabbrev]{cleveref}
\usepackage[toc,page]{appendix}

\usepackage{titlesec}
\usepackage{setspace}

\usepackage{feynman_diagrams}

\crefrangeformat{equation}{(#3#1#4)-(#5#2#6)}
\numberwithin{equation}{section}

\renewcommand{\thesubsection}{\arabic{section}.\arabic{subsection}}

\renewcommand{\L}{\mathcal{L}}

\newcommand{\hf}{\frac{1}{2}}

\newcommand{\abs}[1]{\left|#1\right|}

\DeclareMathOperator{\Tr}{Tr}

\DeclareMathOperator{\sdet}{sdet}

\newdateformat{monthyeardate}{%
  \monthname[\THEMONTH] \THEYEAR}

\begin{document}
\addtocontents{toc}{\protect\setstretch{1.1}}

\hphantom\\
\begin{flushright}
MAN/HEP/2020/004
\\
\monthyeardate{\today}
\end{flushright}

 \title{{\Huge Frame Covariant Formalism for\\[3mm] Fermionic Theories}\\ \vspace{1em} }

\author{\large Kieran Finn}
\author{\large Sotirios Karamitsos}
\author{\large Apostolos Pilaftsis}

\affiliation{\vspace{0.4em}Department of Physics and Astronomy, University of Manchester, Manchester
 M13 9PL, United Kingdom}

\begin{abstract}
\noindent
We present a frame- and reparametrisation-invariant formalism for quantum field theories that include fermionic degrees of freedom. We achieve this using methods of field-space covariance and the Vilkovisky--DeWitt~(VDW) effective action. We explicitly construct a field-space supermanifold on which the quantum fields act as coordinates. We show how to define field-space tensors on this supermanifold from the classical action that are covariant under field reparametrisations. We then employ these tensors to equip the field-space supermanifold with a metric, thus solving a long-standing problem concerning the proper definition of a metric for fermionic theories. With the metric thus defined, we use well-established field-space techniques to extend the VDW effective action and express any fermionic theory in a frame- and field-reparametrisation-invariant manner.\clearpage
\end{abstract}

\maketitle

\section{Introduction}\label{sec:intro}

The same theory of physics can often be written in many different ways. By choosing a different parametrisation of the underlying degrees of freedom, one can make the theory appear very different. However, although intermediate calculations may differ, the predictions of the theory should be the same regardless of which parametrisation is chosen. This idea is known as \emph{reparametrisation invariance}.

Although one might expect reparametrisation invariance to be satisfied almost trivially, there are actually several theories of physics for which it is not obeyed, most notably Quantum Field Theories (QFTs). As was first noted by Vilkovisky~\cite{Vilkovisky:1984st,Vilkovisky:1984un}, the standard definition of the quantum effective action~\cite{Sauter:1931zz,Heisenberg:1935qt,Weisskopf:1996bu,Schwinger:1951nm} for QFTs yields different results off-shell for different parametrisations of the same theory~\cite{Kunstatter:1986qa,Burgess:1987zi,Kunstatter:1990hz,Odintsov:1989gz}. These differences persist even when calculations are performed perturbatively using Feynman diagrams~\cite{Finn:2019aip}.

There have been several attempts to rectify these issues~
 and define a formalism in which reparametrisation invariance is made manifest. Most progress in this direction has been made using the technique of field-space covariance~\cite{Vilkovisky:1984st,Vilkovisky:1984un,Gell-Mann1960,DeWitt:1985sg,Rebhan:1986wp,Ellicott:1987ir,GrootNibbelink:2000vx,GrootNibbelink:2001qt,vanTent:2003mn,Burns:2016ric,Karamitsos:2017elm}. In such an approach, the fields are interpreted as coordinates on a manifold, known as the field-space manifold. In this way, field redefinitions can be interpreted as diffeomorphisms of the field space. With this identification, we can lean on the vast resources of differential geometry to construct theories for which reparametrisation invariance is made manifest. All we need is to work exclusively with field-space tensors and ensure that all field-space indices are fully contracted. This formalism led to the reparametrisation invariant Vilkovisky--DeWitt (VDW) effective action~\cite{Vilkovisky:1984st,Vilkovisky:1984un,DeWitt:1985sg}.

The method of field-space covariance has been very successful for scalar field theories and has also been applied to gravity~\cite{Vilkovisky:1984st,Vilkovisky:1984un,Finn:2019aip,DeWitt:1967ub,DeWitt:1967uc,DeWitt:1967yk,Huggins:1986ht,Lavrov:1988is,Buchbinder:1988np,Steinwachs:2013tr,Moss:2014nya,Bounakis:2017fkv} and gauge theories~\cite{Vilkovisky:1984st,Vilkovisky:1984un,Kunstatter:1986qa,DeWitt:1980jv,DeWitt:1988dq,Fradkin:1983nw}. However the formalism has had less success in theories with fermionic degrees of freedom. While there have been some attempts to construct the VDW effective action for fermionic theories, these have either made no attempt to calculate the metric of the field-space~\cite{DeWitt:1985sg,Rebhan:1986wp} or used a definition specific to the model under consideration~\cite{Helset:2020yio,Fumagalli:2020ody}. As of yet, there has been no method to systematically define the field-space manifold for fermionic theories and this has potentially prevented the use of the VDW formalism from becoming more widespread.

The utility of generalizing the formalism to such theories should be readily apparent. Fermions are an integral part of all realistic quantum field theories, including the Standard Model~\cite{Yang1954,Glashow1961,Weinberg1967}. By constructing the field space for theories with fermionic degrees of freedom, we will complete the formalism and will therefore be able to describe \emph{all} quantum field theories in a way that is manifestly reparametrisation invariant. The goal of this paper is therefore to explicitly construct a field space for fermionic theories and thereby extend the applicability of the VDW effective action.

There are two main distinctions between fermionic fields and bosonic fields that affect how we construct the field space. First is the fact that fermions anti-commute with each other~\cite{Dirac:1928hu,Fermi:1934hr,Schwinger:1969vz}. This requires the introduction of new mathematics to describe them even at the classical level -- namely the introduction of Grassmannian fields~\cite{Grassmann}. To include such anticommuting degrees of freedom in this formalism, we must generalise the field space to a supermanifold~\cite{Berezin:1987wh,berezin1966method,Kostant:1977ab,Batchelor1979,Batchelor:1979ab,Leites_1980,dewitt1992supermanifolds,Rogers:2007zza}. This is a manifold in which some of the coordinates are Grassmannian. We will discuss the implications of this in more detail in Section~\ref{sec:supermanifold}.

In line with the conventions of the literature on supermanifolds, we will be using the prefix `super' in several contexts. For example, we will employ the \emph{superdeterminant}, \emph{supertranspose} and even a notion of \emph{supersymmetric} (these terms will be defined in due course). We wish to emphasise that, despite their names, these terms have nothing to do with Super\-symmetry (SUSY) as it is usually understood in particle\- physics~\cite{Volkov:1973ix,Wess:1974tw,Sohnius:1985qm}. The formalism derived in this paper is applicable to all theories regardless of their underlying symmetry and thus applies equally to both Supersymmetric and non-Supersymmetric theories. We shall distinguish between the two concepts by writing terms related to supermanifolds in lower case. In contrast, when referring to theories with a physical Grassmannian symmetry, we shall always use the term Supersymmetry with a capital S or the acronym SUSY.

The second novelty of fermionic fields is that their equations of motion are only first order. This is in contrast to the equations of motion for bosonic fields, which are of second order and, in the absence of potential terms, constitute the geodesic equation of the field space. This difference arises from the fact that only single derivatives of fermions appear in the Lagrangian. Because of this, a new definition of the field-space metric is required for such theories. In fact, if we simply kept the same definition as we had for scalar field theories~\cite{Finn:2019aip}, we would get a singular metric when fermionic degrees of freedom are included.

The paper is laid out as follows. We start in Section~\ref{sec:scalars} by reviewing the construction of the field space and the VDW effective action for scalar field theories. In Section~\ref{sec:supermanifold} we then review the properties of supermanifolds and highlight the implications of the Grassmannian nature of fermions. In Section~\ref{sec:field space} we
define a supermanifold for QFTs that include fermionic degrees of freedom. In Section~\ref{sec:tensors} we show how field-space tensors on this supermanifold can be extracted from the Lagrangian. In Section~\ref{sec:metric} we show how to equip the field space supermanifold with a metric. We discuss what properties such a metric should possess in order to be consistent with known results and show how a metric with the correct properties can be constructed explicitly from the Lagrangian. For illustration, we discuss a counterexample of a candidate metric with bad properties in Appendix~\ref{sec:covariant metric}. In Section~\ref{sec:effective action} we then combine these ideas with the VDW effective action in order to construct a fully reparametrisation invariant expression for the effective action. We then show some explicit examples of our construction in Section~\ref{sec:examples}, before summarising our findings in Section~\ref{sec:discussion}.

\section{Field-Space Covariance for Scalar Fields}\label{sec:scalars}


The field-space covariant formalism was pioneered by Vilkovisky~\cite{Vilkovisky:1984st,Vilkovisky:1984un} and DeWitt~\cite{DeWitt:1985sg}. In this section we review their construction of the field space for scalar field theories. This will help us identify what aspects of the construction must be altered when considering fermionic degrees of freedom.

We consider a theory of~$N$ scalar fields~$\phi^A$, collectively denoted as~$\bm\phi$ living in a fixed spacetime with metric~$g_{\mu\nu}$. Such a theory is generally described by the Lagrangian
\begin{equation}
\L=\hf k_{AB}(\bm\phi)g^{\mu\nu}\partial_\mu\phi^A\partial_\nu\phi^B-V(\bm\phi)\;,\label{eq:scalar lagrangian}
\end{equation}
where $k_{AB}(\bm\phi)$ is a general model function for the kinetic term and $V(\bm\phi)$ 
represents the scalar potential.
We define the field space to be an~$N$-dimensional manifold with coordinates~$\phi^A$. In so doing, reparametrisations of the fields
\begin{equation}
\phi^A\to\widetilde{\phi}^A(\bm\phi)
\end{equation}
are interpreted as diffeomorphisms of the field space. We can then impose reparametrisation invariance using well-known techniques from differential geometry.

In order to take full advantage of these techniques, we equip the field-space manifold with a metric
\begin{equation}
G_{AB}=\frac{g_{\mu\nu}}{4}\frac{\partial^2\L}{\partial(\partial_\mu\phi^A)\partial(\partial_\nu\phi^B)}-\frac{\partial^2\L}{\partial\phi^A\partial(\square\phi^B)}-\frac{\partial^2\L}{\partial\phi^B\partial(\square\phi^A)},\label{eq:scalar metric}
\end{equation}
which, for the Lagrangian~\eqref{eq:scalar lagrangian}, gives~$G_{AB}=k_{AB}$. Note that this definition differs from the one in~\cite{Finn:2019aip} by the addition of the last two terms. These terms ensure that the metric does not depend on total derivatives appearing in the Lagrangian. The two definitions are identical for Lagrangians that contain no second derivative terms, as it is the
case for~\eqref{eq:scalar lagrangian}.

With the metric defined in~\eqref{eq:scalar metric}, we may introduce a connection into the field space via the Christoffell symbols
\begin{equation}
\Gamma^A_{BC}\equiv\hf G^{AD}\left[\frac{\partial G_{BD}}{\partial \phi^C}+\frac{\partial G_{DC}}{\partial \phi^B}-\frac{\partial G_{BC}}{\partial \phi^D}\right]\,,
\label{eq:scalar field space connection}
\end{equation}
where~$G^{AB}$ is the inverse of~$G_{AB}$. Hence, we are able to define a covariant derivative on the field space:
\begin{align}
     \label{eq:scalar field space cov der}
\nabla_C X^A&=\frac{\partial X^A}{\partial \phi^C}+\Gamma^A_{CD}X^D,&\nabla_C X_A&=\frac{\partial X_A}{\partial \phi^C}-\Gamma^D_{CA}X_D\;,              
\end{align}
with obvious generalisation to higher order tensors.

To take account of the spacetime dependence of the fields, the field-space manifold is often generalised to the infinite-dimensional \emph{configuration space} manifold by taking each spacetime configuration of the fields as a different dimension on the manifold. This manifold can be described by coordinates
\begin{equation}
\phi^{\widehat{A}}\equiv\phi^A(\bm x_A).
\end{equation}
Here we have introduced a condensed notation in which an index with a hat~$\widehat{}$ ~represents both a discrete field-space index and a point in spacetime.

The metric of the configuration space is given by
\begin{equation}
G_{\widehat{A}\widehat{B}} =
\frac{g_{\mu\nu}}{4}\frac{\delta^2S}{\delta(\partial_\mu\phi^{\widehat{A}})\delta(\partial_\nu\phi^{\widehat{B}})}
=G_{AB}\delta^{(4)}(\bm{x}_A-\bm{x}_B),\label{eq:scalar config metric}
\end{equation}
where $S=\int d^4x\sqrt{-g} \L$ is the action and $\delta^{(4)}(\bm{x})$ is the covariant four-dimensional Dirac delta function, normalised such that
\begin{equation}
\int \sqrt{-g}\,\delta^{(4)}(\bm{x})\,d^4x=1.
\end{equation}
The definition of the configuration space metric leads to the following definition for the configuration space connections
\begin{align}
\begin{aligned}
\Gamma^{\widehat{A}}_{\widehat{B}\widehat{C}}&\equiv\hf G^{\widehat{A}\widehat{D}}\left[\frac{\delta G_{\widehat{B}\widehat{D}}}{\delta \phi^{\widehat{C}}}+\frac{\delta G_{\widehat{D}\widehat{C}}}{\delta \phi^{\widehat{B}}}-\frac{\delta G_{\widehat{B}\widehat{C}}}{\delta \phi^{\widehat{D}}}\right]\\
&=\Gamma^A_{BC}\delta^{(4)}(\bm{x}_A-\bm{x}_B)\delta^{(4)}(\bm{x}_A-\bm{x}_C)
\end{aligned}
\label{eq:scalar config space connection}
\end{align}
and hence configuration space covariant derivatives
\begin{align}
\nabla_{\widehat{C}} X^{\widehat{A}}&=\frac{\delta X^{\widehat{A}}}{\delta \phi^{\widehat{C}}}+\Gamma^{\widehat{A}}_{\widehat{C}\widehat{D}}X^{\widehat{D}},&\nabla_{\widehat{C}} X_{\widehat{A}}&=\frac{\delta X_{\widehat{A}}}{\delta \phi^{\widehat{C}}}-\Gamma^{\widehat{D}}_{\widehat{C}\widehat{A}}X_{\widehat{D}},\label{eq:scalar config space cov der}
\end{align}
with straightforward generalisation to higher order tensors. Here we have adopted the condensed Einstein-DeWitt notation~\cite{DeWitt:1967ub} in which repeated configuration space indices imply summation over the discrete index and integration over spacetime.

Having set up the field space and configuration space technology, we may now  use the results of Vilkovisky~\cite{Vilkovisky:1984st,Vilkovisky:1984un} and DeWitt~\cite{DeWitt:1985sg} to write a reparametrisation-invariant expression for the quantum effective action
\begin{align}\label{eq:VDW effective action}
 {\exp\left(\frac{i}{\hbar}   \Gamma[\bm{\varphi}]  \right)} =\int& [\mathcal{D}\bm{\phi}]\, \sqrt{\det G(\bm\phi)}\, \exp \big[ \frac{i}{\hbar}
\Big( S[   \bm{\phi}] +   \; \int d^4x\sqrt{-g}\, \frac{\partial\Gamma[\bm{\varphi}  ]}{\partial\varphi^A} \, \Sigma^A[\bm{\varphi},\bm{\phi}]  \,   \Big)\big].
\end{align}
Here $\bm\varphi$ stands for the mean field and
 \begin{align}
\label{eq:def Sigma}
\Sigma^A [\bm{\varphi},\bm{\phi}] =  (C^{-1}[\bm{\varphi}])^A_{\ B} \, \sigma^B[\bm{\varphi},\bm{\phi}]
\end{align}
is a linear combination of the tangent vectors to the geodesics connecting~$\bm{\varphi}$ and~$\bm{\phi}$, which are denoted~$\sigma^A[\bm{\varphi},\bm{\phi}]$. The matrix~$C[\bm{\varphi}]^A_{\ B}$ is chosen such that that~$\langle\Sigma^A\rangle=0$ and thus all tadpole diagrams evaluate to zero. It can be shown that~$C[\bm{\varphi}]^A_{\ B}$ can be expanded as~\cite{DeWitt:1985sg}
\begin{align}
 \label{eq:expand C}
C ^A_{\ B}[\bm\varphi]=&\Big< \nabla_B\sigma^A[\bm\varphi,\bm\phi]\Big> =\Big< \delta^A_B - \frac{1}{3}  R^A_{\ CBD}[\bm{\varphi}]\ \sigma^C [\bm{\varphi}, \bm{\phi}]\; \sigma^D[\bm{\varphi}, \bm{\phi}] + \ldots\Big>
\end{align}
where~$R^A_{\ CBD}$ is the field-space Riemann tensor and all covariant derivatives are taken with respect to the mean field $\bm\varphi$. Similarly,~$\sigma^A[\bm{\varphi},\bm{\phi}]$ can be expanded as~\cite{Vilkovisky:1984st,Vilkovisky:1984un}
  \begin{align}
 -\sigma^A[\bm{\varphi},\bm{\phi}]  =  -(\varphi^A-\phi^A) + \, \frac{1}{2} \Gamma^A_{BC}[\bm{\varphi}] (\varphi^B-\phi^B)(\varphi^C-\phi^C) \; +\; \cdots.
\end{align}

The VDW effective action can be expanded perturbatively using the background field method to give, at one and two loop orders~\cite{Ellicott:1987ir},
 \begin{align} 
\Gamma^{(1)}  [  \bm{\varphi}]   \; = \;  &-\frac{i}{2}\ln \det   \nabla^{\widehat{A}} \nabla_{\widehat{B}} S, \label{eq:scalar 1 loop}\\
\Gamma^{(2)}  [  \bm{\varphi}]   \; = \;  &\frac{1}{8}\Delta^{\widehat{A}\widehat{B}}\Delta^{\widehat{C}\widehat{D}}\nabla_{(\widehat{A}}\nabla_{\widehat{B}}\nabla_{\widehat{C}}\nabla_{\widehat{D})} S\\&-\frac{1}{12}\Delta^{\widehat{A}\widehat{B}}\Delta^{\widehat{C}\widehat{D}}\Delta^{\widehat{E}\widehat{F}}\times \big(\nabla_{(\widehat{A}}\nabla_{\widehat{C}}\nabla_{\widehat{E})} S\big)\big(\nabla_{(\widehat{B}}\nabla_{\widehat{D}}\nabla_{\widehat{F})} S\big),\nonumber
 \end{align}
respectively, where~$S=S[\bm\varphi]$ is the action expressed in terms of the mean field $\bm\varphi$, ${\Delta^{\widehat{A}\widehat{B}}=\left(\nabla_{\widehat{A}}\nabla_{\widehat{B}} S\right)^{-1}}$ is the covariant propagator, and the parentheses~$(\dots)$ denote symmetrisation with respect to the indices enclosed.

In addition to computing quantum corrections using the effective action formalism, we can also calculate corrections perturbatively using Feynman diagrams~\cite{Feynman1949ab}. In order to maintain reparametrisation invariance, these should be calculated covariantly using covariant Feynman rules
\begin{equation}
\SolidNPoint{A_1}{A_2}{A_3}{A_n}=\nabla_{(\widehat{A}_{1}}\ldots \nabla_{\widehat{A}_n)}S.\label{eq:cov feyn rule}
\end{equation}
Notice that for theories with a non-trivial field space, these covariant Feynman rules differ from the usual ones. This is entirely by design, since the usual Feynman rules are not field-space tensors and so can lead to results that depend on the pararametrisation of the fields~\cite{Finn:2019aip}.

\section{Supermanifolds}\label{sec:supermanifold}

In order to construct a field-space manifold for fermionic theories we must extend the notion of a Riemannian manifold to include anticommuting coordinates. This is the definition of supermanifold~\cite{Berezin:1987wh,berezin1966method,Kostant:1977ab,Batchelor1979,Batchelor:1979ab,Leites_1980}. In this section we review the basic properties of supermanifolds that will affect our construction of the field space. We encourage interested readers to consult~\cite{dewitt1992supermanifolds,Rogers:2007zza} for further details of the rich mathematics of this topic.

Originally, supermanifolds were invented in the context of Supersymmetry (SUSY)~\cite{Salam:1974ab,Gates:1983nr}. In this context, the usual spacetime is augmented with new Grassmannian coordinates and diffeomorphisms of the new \emph{superspace} result in Grassmannian Noether symmetries. However, since their invention, the mathematics of supermanifolds has been developed as a subject in its own right and now has applications far beyond SUSY~\cite{Guajardo:2014ab,Khudaverdian:2020ab,Kellett:2020rjw}.

It is in this latter context that we employ supermanifolds in this paper. We are extending not the spacetime manifold, but the field-space manifold. Thus, the new Grassmannian diffeomorphisms are not physical Grassmannian symmetries of the theory, but merely reparametrisations of the fermionic fields. As with all reparametrisations, these field-space diffeomorphisms cannot be considered a symmetry in the traditional sense and there will, in general, be no Noether current or gauge degrees of freedom associated with them.


To set our notation, we consider a supermanifold with~$n$ commutative coordinates and~$m$ anti-commutative coordinates. We denote the coordinates~$x^\alpha$ with ${\alpha=(1,2\ldots m+n)}$. When we need to refer to the commutative and anti-commutative coordinates separately we shall use~$x^A$ for the former and~$x^I$ for the latter, with letters from the start of the Latin alphabet indicating commutative coordinates and letters from the middle of the same alphabet indicating anti-commutative coordinates.

The first subtlety we must consider is that when differentiating with respect to an anticommuting coordinate, we must specify whether we are differentiating from the left or from the right. The two types of differentiation are related by
\begin{equation}
\overrightarrow{\partial}_{\!\!\alpha}X=(-1)^{\alpha(X+1)}\;X\overleftarrow{\partial}_{\!\!\alpha}.~\label{eq:left right derivative}
\end{equation}
In the above, we have introduced a new notation common throughout the literature on supermanifolds. The expressions in an exponent of~$-1$ are not meant to be taken literally, but as labels standing for the grading of their respective quantities:~$1$ for anticommuting quantities and~$0$ for commuting quantities. Thus, the quantity~$\alpha$ that appears in the prefactor
on the RHS of~\eqref{eq:left right derivative} is not to be considered an index and is not summed over as would be expected by the Einstein summation convention. Instead, it should be regarded as a label that is~$1$ when~$\alpha$ refers to an anticommuting coordinate and~$0$ when~$\alpha$ refers to a commuting coordinate. Similarly, the~$X$ in the exponent of~$-1$ in~\eqref{eq:left right derivative} is to be considered a label that is~$1$ when~$X$ is an anticommuting object and~$0$ when~$X$ is a commuting object. Thus, \eqref{eq:left right derivative} tells us that there is a factor of~$-1$ between a left and right derivative, when differentiating a commuting object with respect to an anticommuting coordinate, but that they are identical in all other cases.

When performing a diffeomorphism
\begin{equation}
x^\alpha\to\widetilde{x}^\alpha=\widetilde{x}^\alpha(\bm x),\label{eq:superman diffeomorphism}
\end{equation}
this difference between left and right derivatives leads to a distinction between left and right Jacobians. These are, respectively,
\begin{align}
{}_\alpha J^\beta&=\frac{\overrightarrow{\partial}}{\partial x^\alpha}\widetilde{x}^\beta, &{}^\beta J^{\sf{s\!T}}_\alpha&=\widetilde{x}^\beta \frac{\overleftarrow{\partial}}{\partial x^\alpha}.
\end{align}
We distinguish between tensors that transform with a left or right Jacobian by writing the appropriate index to the right or left, respectively, of the tensor. Thus, $V^\alpha$ is a vector that transforms with a left Jacobian and~${}^\alpha V$ is a vector that transforms with a right Jacobian,
e.g.~$\widetilde{V}^\alpha = V^\beta {}_\beta J^\alpha$ and ${}^\alpha \widetilde{V} = {}^\alpha J^{\sf{s\!T}}_\beta\, {}^\beta V$. Similarly, we define left and right covectors~$V_\alpha$ and~${}_\alpha V$ that transform with the left and right inverse Jacobians, respectively.

The superscript~${\sf s\!T}$ denotes the operation of \emph{supertransposition} and is defined as
\begin{align}
\begin{aligned}
{}^\alpha M^{\sf s\!T}_\beta&=(-1)^{\beta(\alpha+1)}\;{}_\beta M^\alpha,&
{}_\alpha M^{\sf s\!T}_\beta&=(-1)^{\alpha+\beta+\alpha\beta}\;{}_\beta M_\alpha,&
{}^\alpha M^{\beta\,{\sf s\!T}}&=(-1)^{\alpha\beta}\;{}^\beta M^\alpha.
\end{aligned}
\end{align}
Note that the rules of supertransposition are different depending on the index placement. The supertranspose satisfies the identities one would expect, namely
\begin{align}
\begin{aligned}
(M^{\sf s\!T})^{\sf s\!T}&=M,&
(M^{-1})^{\sf s\!T}&=(M^{\sf s\!T})^{-1},&
(MN)^{\sf s\!T}&=N^{\sf s\!T} M^{\sf s\!T}.
\end{aligned}
\end{align}
Note that this is in contrast to the regular transpose for which~$(MN)^{\sf T}\neq N^{\sf T} M^{\sf T}$ in the presence of anticommuting coordinates. 

The definition of the supertranspose leads to the notion of supersymmetric and anti-supersymmetric matrices, which satisfy~$M^{\sf s\!T}=M$ and~$M^{\sf s\!T}=-M$, respectively. Again, we emphasise that the definition of supersymmetric here should not be confused with the theory of SUSY.

Finally, we consider the \emph{superdeterminant}, which is sometimes known as the Berezinian~\cite{berezin1966method}. To define the superdeterminant, we consider a square rank-2 tensor on the supermanifold (sometimes known as a supermatrix), which has the form
\begin{equation}
{}_\alpha M_\beta=\begin{pmatrix}
{}_AA_B&{}_AC_J\\
{}_ID_B&{}_IB_J
\end{pmatrix}.
\end{equation}
Here~${}_AA_B$ and ~${}_IB_J$ are~$n\times n$ and~$m\times m$ matrices of commuting numbers, respectively and~${}_AC_J$ and~${}_ID_B$ are~$n\times m$ and~$m\times n$ matrices of anticommuting numbers, respectively. The superdeterminant of such a matrix is given by
\begin{equation}
\sdet{M}=\frac{\det(A-CB^{-1} D)}{\det B}.
\end{equation}
The superdeterminant, defined in this way, is such that the Berezinian integral measure
\begin{equation}
\sqrt{\sdet(M)}\;d^{n+m}x
\end{equation}
is invariant under diffeomorphisms of the supermanifold~\eqref{eq:superman diffeomorphism}~\cite{berezin1966method}. This is true for any rank-2 tensor~${}_\alpha M_\beta$.

\section{The Field Space for Scalar-Fermion Theories}\label{sec:field space}

In this section we construct the field space for a theory with~$N$ real scalar fields and~$M$ Dirac fermions.\footnote{A generalisation of this approach to theories with Weyl fermions will be straightforward.} Recalling that, in 4-dimensional spacetime, each Dirac fermion propagates four complex, or eight real, anticommuting degrees of freedom~\cite{Dirac:1928hu}, we see that the field space for such a theory should be a supermanifold with~$N$ commuting coordinates and~$8M$ anticommuting coordinates. Throughout this paper, we shall use the following set of coordinates to describe this field-space supermanifold
\begin{equation}
\Phi^\alpha=\left(\phi^A\!,\;\psi^1_a,\;\overline{\psi}{}^1_{\dot{a}},\;\psi^2_a,\;\overline{\psi}{}^2_{\dot{a}},\;\ldots\right),
\end{equation}
where the subscripts~$a$ and~$\dot{a}$ refer to the spinor components of the Dirac fermions. As done previously, we use Greek indices from the beginning of the alphabet, which run~${1\leq\alpha\leq N+8M}$, for the full supermanifold and, when we need to refer to them separately, we will use capital Latin letters from the beginning of the alphabet for the commuting coordinates and capital Latin letters from the middle of the alphabet for anticommuting coordinates. These run~${1\leq A\leq N}$ and~${1\leq I\leq 8M}$, respectively. We will refer to all the field-space coordinates collectively as~$\bm\Phi$.

A general field reparametrisation of the form
\begin{equation}
\Phi^\alpha\to\widetilde{\Phi}^\alpha=\widetilde{\Phi}^\alpha(\bm\Phi)\label{eq:field redef}
\end{equation}
is equivalent to a diffeomorphism of the field-space supermanifold and we can therefore enforce reparametrisation invariance using the techniques of differential supergeometry discussed in Section~\ref{sec:supermanifold}. We note that the transformation~\eqref{eq:field redef} is quite general and can, for instance, involve Grassmannian parameters. However, the transformation cannot depend on derivatives of the fields and so there are still certain transformations that are not captured by~\eqref{eq:field redef}, for example the transformations of SUSY~\cite{Volkov:1973ix,Wess:1974tw}. We believe extending the formalism to include such transformations will not be too difficult, but choose to leave such considerations for future work.

With the field space thus defined, let us write the most general Lagrangian for this theory using field-space tensors. Including terms up to quadratic order in derivatives, we get
\begin{equation}
\L=\hf g^{\mu\nu} \partial_\mu \Phi^\alpha\;{}_\alpha k_\beta(\bm\Phi)\;\partial_\nu\Phi^\beta+\frac{i}{2}\zeta_\alpha^\mu(\bm\Phi)\;\partial_\mu\Phi^\alpha-U(\bm\Phi).\label{eq:tensor lagrangian}
\end{equation}
This expression contains three model functions that define our theory:~${}_\alpha k_\beta(\bm\Phi)$ is a rank-2 field-space tensor,~$\zeta_\alpha^\mu(\bm\Phi)$ is a field-space covector and a spacetime vector and $U(\bm\Phi)$ is a field-space and spacetime scalar. These model functions can, in general, depend on  both the scalar and fermion fields in the theory, but not on their derivatives.

Let us analyse these three model functions in turn. The tensor~${}_\alpha k_\beta$ is the scalar field-space metric. In the absence of fermion fields, this object just reduces to the metric~\eqref{eq:scalar metric}. Because fermionic fields only enter the Lagrangian with a single derivative, we see that this tensor only has support in the bosonic sector, implying that
\begin{equation}
{}_\alpha k_I={}_I k_\alpha=0.
\end{equation}
Because of this, the tensor~${}_\alpha k_\beta$ is singular and cannot play the role of the field-space metric in the presence of fermions.

Next we look at the potential term~$U(\bm\Phi)$. This term contains the scalar potential~$V(\bm\phi)$, as well as both the fermion mass terms and any momentum-independent interactions between the scalars and fermions such as Yukawa interactions~\cite{Yukawa:1935xg}. As in the scalar case, the potential~$U(\bm\Phi)$ plays no role in the construction of the field-space manifold and so acts only as an external force.

Finally, we consider the model function~$\zeta_\alpha^\mu$. This model function has no analogue in a pure scalar field theory, because scalars cannot appear in the Lagrangian with a single derivative in a diffeomorphism invariant way. Since~$\zeta_\alpha^\mu$ cannot depend on derivatives of the fields and there are no spacetime vectors in this theory, the spacetime index~$\mu$ of this tensor can only come from a~$\gamma^\mu$ matrix. This tells us why such a term cannot appear in a pure scalar  theory, since there would then be no fermions to contract with the spinor indices of~$\gamma^\mu$. 

As an example to show the ubiquity of the expression~\eqref{eq:tensor lagrangian}, we consider a theory of free scalars and fermions with Lagrangian
\begin{align}
\begin{aligned}
\L=&\sum_{A\in\mathrm{scalars}}\left[\hf  g^{\mu\nu}\partial_\mu\phi^A\partial_\nu\phi^A-\hf m_A^2(\phi^A)^2\right]\\
&+\sum_{X\in\mathrm{fermions}}\Big[\frac{i}{2}\left(\overline{\psi}{}^X\gamma^\mu\partial_\mu\psi^X-\partial_\mu\overline{\psi}{}^X\gamma^\mu\psi^X\right)-m_X\overline{\psi}^X\psi^X\Big]. 
\end{aligned}\label{eq:free lagrangian}
\end{align}
Such a theory has the following model functions:
\begin{align}
\begin{aligned}
{}_\alpha k_\beta&=\begin{pmatrix}
\delta_{AB}&\bm0_{N\times8M}\\
\bm0_{8M\times N}&\bm0_{8M\times 8M}
\end{pmatrix},\\
\zeta_\alpha^\mu&=\left({\bm 0}_N,\;\overline{\psi}{}^1_{\dot{a}}\gamma^\mu_{\dot{a}a},\;\gamma^\mu_{\dot{a}a}\psi^1_a,\;\overline{\psi}{}^2_{\dot{b}}\gamma^\mu_{\dot{b}b},\;\gamma^\mu_{\dot{b}b}\psi^2_b,\;\ldots\right)\\
U&=\sum_{A\in\mathrm{scalars}}\hf m_A^2(\phi^A)^2+\sum_{X\in\mathrm{fermions}}m_X\overline{\psi}^X\psi^X.
\end{aligned}
\end{align}
We will discuss more general examples in Section~\ref{sec:examples}.

We can extract the model functions from the Lagrangian in a constructive manner
with the following definitions
\begin{align}
{}_\alpha k_\beta&=\frac{g_{\mu\nu}}{4}\frac{\overrightarrow{\partial}}{\partial(\partial_\mu\Phi^\alpha)}\;\L\;\frac{\overleftarrow{\partial}}{\partial(\partial_\nu\Phi^\beta)},\label{eq:extract k}\\
\zeta_\alpha^\mu&=\frac{2}{i}\left(\L-\hf g^{\mu\nu} \partial_\mu \Phi^\alpha\;{}_\alpha k_\beta\;\partial_\nu\Phi^\beta\right)\frac{\overleftarrow{\partial}}{\partial(\partial_\mu\Phi^\alpha)}\label{eq:extract zeta},\\
U&=\hf g^{\mu\nu} \partial_\mu \Phi^\alpha\;{}_\alpha k_\beta(\bm\Phi)\;\partial_\nu\Phi^\beta+\frac{i}{2}\zeta_\alpha^\mu(\bm\Phi)\;\partial_\mu\Phi^\alpha-\L.
\end{align}
Such a construction is important in ensuring the formalism developed in this paper is unique.

\section{Tensors in the Field Space}\label{sec:tensors}

Having defined the field-space supermanifold, we now wish to investigate which field-space tensors can be constructed, given only the Lagrangian of the theory. The ultimate goal is to define a metric for the field space so that we can apply the formalism of the VDW effective action.

The model function~$\zeta_\alpha^\mu$ is a covector in field space, but it is also a spacetime vector. It would be nice to remove the~$\mu$ index somehow in order to obtain a pure field-space covector. The simplest way to achieve this would be to contract~$\zeta_\alpha^\mu$ with some spacetime covector. However, there are no spacetime covectors in the theory and~$\zeta_\alpha^\mu$ does not have the right spinor structure to allow a contraction with a~$\gamma^\mu$ matrix.

Instead, we rely on the observation noted earlier -- the spacetime properties of~$\zeta_\alpha^\mu$ are inherited from a~$\gamma^\mu$ matrix. We can therefore render a pure field-space covector from~$\zeta_\alpha^\mu$ by surgically removing the~$\gamma^\mu$ matrices. 

We can do this in a rigorous way by defining the notion of differentiation with respect to a $\gamma^\mu$ matrix. In order to define such a notion, we remember that any matrix $M_{a\dot{a}}$ in spinor space can be uniquely expressed in terms of 16 orthogonal Lorentz-covariant bilinears as
\begin{align}
   \label{eq:generalspinormatrix}
\begin{aligned}
M_{a\dot{a}}&=\sum_{i=S,P,V,A,T} a^{(i)} \;\Gamma^{(i)}_{a\dot{a}},& \Gamma^{(i)}\in &\ \{I_4,\gamma^5,\gamma^\mu,\gamma^\mu\gamma^5,\sigma^{\mu\nu}\}\,,
\end{aligned}
\end{align}
where $\Gamma^{(i)}$ (with $i = S,P,V,A,T$) are the basis matrices, $a^{(i)}$ is a coefficient, and ${\sigma^{\mu\nu}\equiv i/2\,[\gamma^\mu,\gamma^\nu]}$. We can use this expansion to define the
general partial differentiation,
\begin{equation}
   \label{eq:diffgamma}
\frac{\delta F}{\delta\Gamma^{(i)}}\ \equiv\ \lim_{\epsilon^{(i)}\to0} \frac{F[\Gamma^{(i)}\to \Gamma^{(i)} + \epsilon^{(i)} I_4]-F[\Gamma^{(i)}]}{\epsilon^{(i)}}\ ,
\end{equation}
such that when applied to a general matrix $M_{a\dot{a}}$ for $\Gamma^{(V)} = \gamma^\mu$, 
we have
\begin{equation}
\frac{\delta M_{a\dot{a}}}{\delta\gamma^\mu}=a^{(V)}_\mu\,\delta_{a\dot{a}}\; .
\end{equation}
Note that we must insist that all spinor matrices are written in the form~\eqref{eq:generalspinormatrix} before applying~\eqref{eq:diffgamma} to ensure that no ambiguities arise from the Clifford algebra.

We may therefore define\footnote{The factor of 1/4 is included in order to compensate for the factor of 4 arising from the contraction of spacetime indices.}
\begin{equation}
\zeta_\alpha=\frac{1}{4}\,\frac{\delta \zeta^\mu_\alpha}{\delta\gamma^\mu}\; .\label{eq:def zeta}
\end{equation}
For the free theory~\eqref{eq:free lagrangian} this gives
\begin{equation}
\zeta_\alpha=\left({\bm 0}_N,\;\overline{\psi}{}^1_{\dot{a}},\;\psi^1_a,\;\overline{\psi}{}^2_{\dot{b}},\;\psi^2_b,\;\ldots\right).
\end{equation}

The quantity~$\zeta_\alpha$ defined in~\eqref{eq:def zeta} is a true field-space covector and transforms as
\begin{equation}
\zeta_\alpha\to\widetilde{\zeta}_\alpha=\zeta_\beta\;\,{}^\beta\! \left(J^{-1}\right)^{\sf s\!T}_\alpha
\end{equation}
under a field redefinition~\eqref{eq:field redef}, where
\begin{equation}
{}^\beta\! \left(J^{-1}\right)^{\sf s\!T}_\alpha= \Phi^\beta\frac{\overleftarrow{\partial}}{\partial \widetilde{\Phi}^\alpha}
\end{equation}
is the right inverse Jacobian of the transformation.

From~$\zeta_\alpha$, we can also define a rank-2 tensor
\begin{equation}
{}_\alpha \lambda_\beta=\frac{\overrightarrow{\partial}}{\partial\Phi^\alpha}\zeta_\beta-(-1)^{\alpha+\beta+\alpha\beta}\frac{\overrightarrow{\partial}}{\partial\Phi^\beta}\zeta_\alpha\,.\label{eq:define lambda}
\end{equation}
Despite the appearance of partial non-covariant derivatives in~\eqref{eq:define lambda},~${}_\alpha \lambda_\beta$ is still a rank-2 tensor and transforms as
\begin{equation}
{}_\alpha \lambda_\beta\to{}_\alpha\widetilde{\lambda}_\beta={}_\alpha \!\left(J^{-1}\right)^\gamma\;{}_\gamma \lambda_\delta\;\,{}^\delta \!\left(J^{-1}\right)^{\sf s\!T}_\beta.
\end{equation}
under a field redefinition~\eqref{eq:field redef}. The reason why~${}_\alpha \lambda_\beta$ transforms as a tensor despite the use of partial derivatives is the same reason that the field strength tensor~$F_{\mu\nu}$ transforms as a spacetime tensor in QED -- namely, any connections added to~\eqref{eq:define lambda} would cancel between the two terms. This cancellation also ensures that~${}_\alpha \lambda_\beta$ is left unchanged by the addition of a total derivative to the Lagrangian.  Note that the matrix~${}_\alpha \lambda_\beta$ is odd under supertransposition (i.e. it is anti-supersymmetric) obeying the property~$\lambda^{\sf s\!T}=-\lambda$.

Because of the presence of the scalar fields, the matrix~${}_\alpha \lambda_\beta$ is singular even for well behaved theories limiting its usefulness. However, the sum
\begin{equation}
{}_\alpha \Lambda_\beta={}_\alpha k_\beta+{}_\alpha \lambda_\beta\label{eq:define Lambda}
\end{equation}
is non-singular and therefore has an inverse and a non-zero superdeterminant.

For the free theory~\eqref{eq:free lagrangian}, the definition~\eqref{eq:define Lambda} gives
the metric
\begin{equation}
{}_\alpha \Lambda_\beta={}_\alpha N_\beta\equiv\begin{pmatrix}
1_N&0&0&0&0&\cdots\\
0&0&1_4&0&0&\cdots\\
0&1_4&0&0&0&\cdots\\
0&0&0&0&1_4&\cdots\\
0&0&0&1_4&0&\cdots\\
\vdots&\vdots&\vdots&\vdots&\vdots&\ddots
\end{pmatrix}.\label{eq:flat lambda}
\end{equation}

\section{The Field-Space Metric}\label{sec:metric}

We now wish to define a metric for the field-space supermanifold. Such a metric should satisfy the following properties:
\begin{enumerate}
\item The metric should be determined solely and uniquely from the action.
As a corollary of this property, total derivatives should not contribute to the metric.\label{prop:unique}
\item The metric should transform as a rank 2 field-space tensor.\label{prop:tensor}
\item The metric should be supersymmetric, as any antisupersymmetric part will not contribute to the line element of the field-space supermanifold.\label{prop:supersym}
\item The metric should not be singular, unless there are non-dynamical degrees of freedom.\label{prop:singular}
\item The metric should depend on the fields only and not on their derivatives.\label{prop:finsler}
\item The metric for a theory with canonically normalised fields should be given by
\begin{equation}
{}_a H_b\equiv\begin{pmatrix}
1_N&0&0&0&0&\cdots\\
0&0&1_4&0&0&\cdots\\
0&-1_4&0&0&0&\cdots\\
0&0&0&0&1_4&\cdots\\
0&0&0&-1_4&0&\cdots\\
\vdots&\vdots&\vdots&\vdots&\vdots&\ddots
\end{pmatrix},\label{eq:flat metric}
\end{equation}
which is the analogue of the Euclidean metric that governs locally the supermanifold~\cite{dewitt1992supermanifolds}.\label{prop:flat}
\end{enumerate}

The tensor~${}_\alpha \Lambda_\beta$ may seem to be a good candidate for the metric, but it does not satisfy property~\ref{prop:supersym}, since ${\Lambda^{\sf s\!T}\neq\Lambda}$. This property is needed for a metric, since only the supersymmetric part of a metric contributes to the field-space line element.  The supersymmetric part of~${}_\alpha \Lambda_\beta$ is~${}_\alpha k_\beta$ which, as we have argued before, cannot be used as a metric because it is singular and so violates property~\ref{prop:singular}.

One might be tempted to alter the definition~\eqref{eq:define lambda} so that there is a relative plus sign between the two terms instead of a minus. However, as we show in Appendix~\ref{sec:covariant metric}, such a definition crucially depends on possible total derivatives that one could add to the Lagrangian and so it violates property~\ref{prop:unique}.

Instead we make use of a useful property of super\-manifolds, which is inherited from their relation to ordinary Riemannian manifolds. A~supermanifold can always be made locally flat by a suitable change of coordinates~\cite{riemann_1851}. Therefore, by switching to these local inertial coordinates, we can render the field-space metric into a known, simple form.

Mathematically, this is expressed in terms of vielbeins~\cite{Schwinger:1963re} as follows:
\begin{equation}
{}_\alpha G_\beta={}_\alpha e^a \; {}_a H_b\;{}^b e^{\sf s\!T}_\beta,\label{eq:metric vielbein}
\end{equation}
where~${}_\alpha G_\beta$ is the field-space metric,~${}_\alpha e^a$ are the vielbeins and ${}_aH_b$ is defined in~\eqref{eq:flat metric}.

But, we know from property~\ref{prop:flat} that in the local inertial frame the fields should be locally canonical with the same kinetic terms as~\eqref{eq:free lagrangian}. Therefore, in the local inertial frame, we must have~${{}_a\Lambda_b={}_aN_b}$ as shown in~\eqref{eq:flat lambda} and so
\begin{equation}
{}_\alpha \Lambda_\beta={}_\alpha e^a \; {}_a N_b\;{}^b e^{\sf s\!T}_\beta.\label{eq:find vielbeins}
\end{equation}
The only unknowns in~\eqref{eq:find vielbeins} are the vielbeins and thus we can use this equation to calculate ${}_\alpha e^a$ and, hence, the field-space metric ${}_\alpha G_\beta$.

The relation~\eqref{eq:find vielbeins} does not define the vielbeins~${}_\alpha e^a$ uniquely, but only up to a matrix~${}_aX^b$ that satisfies
\begin{equation}
{}_a X^c\;{}_c N_d\;{}^d X^{\sf s\!T}_b={}_aN_b\,.
\end{equation}
The most general such matrix can be written as a product of two other matrices as
\begin{equation}
   \label{eq:aXb}
{}_aX^b={}_a Y^c\;{}_cX_0^b\,.
\end{equation}
The second matrix ${}_cX_0^b$ is given by
\begin{equation}
{}_aX_0^b=\begin{pmatrix}
O_N&0&0&0&0&\cdots\\
0&x_1&0&0&0&\cdots\\
0&0&x_1^{-1}&0&0&\cdots\\
0&0&0&x_2&0&\cdots\\
0&0&0&0&x_2^{-1}&\cdots\\
\vdots&\vdots&\vdots&\vdots&\vdots&\ddots
\end{pmatrix},\label{eq:X0}
\end{equation}
where~$O_N$ is an orthogonal~$N\times N$ matrix and~$x_i$ (with $i=1,2,\dots, M$) are a set of~$M$ arbitrary invertible~$4\times 4$ matrices.

The first matrix~${}_aY^b$ in~\eqref{eq:aXb} accounts for the fact that~${}_aN_b$ is invariant under the exchange of~$\psi^I\leftrightarrow \overline{\psi}{}^I$ for any fermion in the theory. Thus, we can multiply~${}_\alpha e^a$ by any matrix that implements such an exchange. There are~$2^M$ such matrices, which have the form 
\begin{equation} 
{}_aY^b=\begin{pmatrix}
    1_N&0&0&\cdots&0\\
    0&y_1&0&\cdots&0\\
    0&0&y_2&\cdots&0\\
    \vdots&\vdots&\vdots&\ddots&\vdots\\
    0&0&0&\cdots& y_M \end{pmatrix},\label{eq:Y} 
\end{equation} 
where each of the~$y_i$ (with $i=1,2,\ldots M$) is equal to either
\begin{equation} \begin{pmatrix}
    1_4&0\\
    0&1_4 \end{pmatrix}\;\;\mathrm{or}\;\;\begin{pmatrix}
    0&1_4\\
    1_4&0 \end{pmatrix}.  \end{equation}

We notice that~${}_a X_0^c\;{}_c H_d\;{}^dX^{\sf s\!T}_{0\,b}={}_aH_b$. Consequently, the choice of~$O_N$ and~$x_i$ does not affect the field-space metric~\eqref{eq:metric vielbein}. However,~${}_a Y^c\;{}_c H_d\;{}^dY^{\sf s\!T}_{b}\neq{}_aH_b$ and hence, the choice of~$y_i$ will result in~$2^M$ different possible metrics.

However, only one of these metrics will have a signature compatible with~\eqref{eq:flat metric} with all minus signs on the lower left diagonal. The other~$2^M-1$ choices of metric will all have at least one minus sign on the upper right diagonal. Therefore, only one choice of~${}_aY^b$ will result in an acceptable field space vielbein. Any other choice would result in a difference between the signature of the local metric and the global metric. Such a signature difference is inadmissible on a supermanifold, in the same way as it is on a regular manifold.

We can, therefore, define a unique field-space metric by insisting that the~$\psi^I\overline{\psi}{}^I$ components of the metric are positive, while the~$\overline{\psi}{}^I\psi^I$ components are negative.\footnote{Note that such a definition can only be made unambiguously if these entries have the same sign everywhere in field-space. This is always the case. For these entries to change sign, they would have to pass through zero, at which point the metric would become singular and the corresponding field would become non-dynamical. Such a situation is unphysical and thus this requirement is unambiguous.}

With the field-space metric ${}_\alpha G_\beta$ determined as described above, we can now proceed to 
evaluate the field-space connections through the Christoffell symbols
\begin{align}
\begin{aligned}
{}^\alpha\Gamma_{\beta\gamma}=\hf{}^\alpha G^\delta\Big[&{}_\delta G_\beta \overleftarrow{\partial}_{\!\gamma}+(-1)^{\beta\gamma}\; {}_\delta G_\gamma\overleftarrow{\partial}_{\!\beta}-(-1)^{\beta}\overrightarrow{\partial}_{\!\delta}\;{}_\beta G_\gamma\Big].
\end{aligned}\label{eq:connections}
\end{align}
We can then use these connections to define covariant derivatives on the field space
\begin{align}
\begin{aligned}
X^\alpha\overleftarrow{\nabla}_{\!\!\beta}&=X^\alpha\frac{\overleftarrow{\partial}}{\partial \Phi^\beta}+{}^\alpha\Gamma_{\beta\gamma}X^\gamma,\\
X_\alpha\overleftarrow{\nabla}_{\!\!\beta}&=X_\alpha\frac{\overleftarrow{\partial}}{\partial \Phi^\beta}-X_\gamma\;{}^\gamma\Gamma_{\alpha\beta},
\end{aligned}
\end{align}
with straightforward generalisation to higher order tensors.

The field-space supermanifold can be straightforwardly generalised to an infinite-dimensional configuration-space manifold with coordinates
\begin{equation}
\Phi^{\widehat{\alpha}}\equiv\Phi^\alpha(\bm{x}_\alpha).\label{eq:config space coords}
\end{equation}
We can define the configuration space metric analogously to~\eqref{eq:scalar config metric} by
\begin{equation}
{}_{\widehat{\alpha}}G_{\widehat{\beta}}={}_\alpha G_\beta \delta^{(4)}(\bm{x}_\alpha-\bm{x}_\beta).
\end{equation}
Similarly, we can define the configuration space connections
\begin{align}
\begin{aligned}
{}^{\widehat{\alpha}}\Gamma_{\widehat{\beta}\widehat{\gamma}}\equiv&\hf {}^{\widehat{\alpha}}G^{\widehat{\delta}}\Big[{}_{\widehat{\delta}}G_{\widehat{\beta}}\frac{\overleftarrow{\delta}}{\delta \phi^{\widehat{\gamma}}}+(-1)^{\beta\gamma}\;{}_{\widehat{\delta}}G_{\widehat{\gamma}}\frac{\overleftarrow{\delta}}{\delta \phi^{\widehat{\beta}}}-(-1)^\beta\frac{\overrightarrow{\delta}}{\delta \phi^{\widehat{\delta}}}\,{}_{\widehat{\beta}}G_{\widehat{\gamma}}\Big]\\
=&{}^\alpha\Gamma_{\beta\gamma}\delta^{(4)}(\bm{x}_\alpha-\bm{x}_\beta)\delta^{(4)}(\bm{x}_\alpha-\bm{x}_\gamma)
\end{aligned}\label{eq:config space connection}
\end{align}
and hence configuration space covariant derivatives
\begin{align}
\begin{aligned}
 X^{\widehat{\alpha}}\overleftarrow{\nabla}_{\widehat{\beta}}&=X^{\widehat{\alpha}}\frac{\overleftarrow{\delta}}{\delta \phi^{\widehat{\beta}}}+{}^{\widehat{\alpha}}\Gamma_{\widehat{\beta}\widehat{\gamma}}X^{\widehat{\gamma}},&
  X_{\widehat{\alpha}}\overleftarrow{\nabla}_{\widehat{\beta}}&=X_{\widehat{\alpha}}\frac{\overleftarrow{\delta}}{\delta \phi^{\widehat{\beta}}}-X_{\widehat{\gamma}}\;{}^{\widehat{\gamma}}\Gamma_{\widehat{\beta}\widehat{\alpha}},
 \end{aligned}\label{eq:config space cov der}
\end{align}
with straightforward generalisation to higher order tensors.

Finally, we can define a reparametrisation invariant measure
\begin{equation}
\left[D{\cal M}\right]=\sqrt{\abs{\sdet G}}\;\left[D^{N+8M}\Phi_q\right],\label{eq:metric measure}
\end{equation}
where $\bm{\Phi}_q$ represent collectively the quantum field coordinates corresponding to $\bm\Phi$ defined in~\eqref{eq:config space coords}. We note that in the local inertial frame~$\abs{\sdet{H}}=\abs{\sdet{N}}$ as can be seen from~\eqref{eq:flat metric} and~\eqref{eq:flat lambda}. Since these two quantities transform in the same way, we therefore conclude that~${\abs{\sdet{G}}=\abs{\sdet{\Lambda}}}$ in all frames. Thus, we can always replace~$G$ with~$\Lambda$ in~\eqref{eq:metric measure}, without affecting the measure if this proves an easier quantity to calculate.

\section{The Covariant Effective Action}\label{sec:effective action}

With the metric, connections and invariant volume element determined for the field space in the previous section,  we are now in a position to make use of the VDW formalism~\cite{Vilkovisky:1984st,Vilkovisky:1984un,DeWitt:1985sg,Rebhan:1986wp,Ellicott:1987ir} and define the quantum effective action for theories with fermionic degrees of freedom in a reparametrisation-invariant manner. This leads us to the following implicit equation:
\begin{align}
\exp(i\Gamma[\bm \Phi])=\int& \sqrt{\abs{\sdet G}}\left[D\bm\Phi_q\right] \exp\Big(iS[\bm\Phi_q]-i\int d^4x\sqrt{-g}\,\Gamma[\bm\Phi]\frac{\overleftarrow{\partial} }{\partial \Phi^\alpha}\Sigma^\alpha[\bm\Phi,\bm\Phi_q]\Big),\label{eq:covariant effective action}
\end{align}
where $\bm\Phi$ denotes the mean field in this section. In the above,~$\Sigma^\alpha[\bm\Phi,\bm\Phi_q]$ is related to the supergeodesic tangent vector~$\sigma^\alpha[\bm\Phi,\bm\Phi_q]$ by
 \begin{align}
\Sigma^\alpha [\bm{\Phi},\bm{\Phi}_q] =  (C^{-1}[\bm{\Phi}])^\alpha_{\ \beta} \, \sigma^\beta[\bm{\Phi},\bm{\Phi}_q]\,,
\end{align}
where~$C^\alpha_{\ \beta}[\bm{\Phi}]$ ensures that tadpoles evaluate to zero.
In~\cite{DeWitt:1985sg}, it was found that
\begin{align}
&C^\alpha_{\ \beta}[\bm{\Phi}]=\Big< \sigma^\alpha[\bm\Phi,\bm\Phi_q]\overleftarrow{\nabla}_{\!\beta}\Big>= \Big< \delta^\alpha_{\ \beta}   -(-1)^{\gamma (\beta +\delta)} \frac{1}{3}R^\alpha_{\ \gamma\beta\delta}[\bm{\Phi}]\  \sigma^\gamma[\bm{\Phi}, \bm{\Phi}_q]\;\sigma^\delta [\bm{\Phi}, \bm{\Phi}_q] + \ldots\Big>\,,\label{eq:choose C}
\end{align}
where the quantum expectation~$\langle\,\rangle$ is calculated using~\eqref{eq:covariant effective action} and the field-space Riemann${}$ tensor $R^\alpha_{\ \gamma\beta\delta}$ characterising the curvature of the supermanifold is defined later in~\eqref{eq:Riemann}. In this VDW formulation, the tangent vector~$\sigma^\alpha[\bm{\Phi},\bm{\Phi}_q]~$ can be expanded to give
\begin{align}
\begin{aligned}
 -\sigma^\alpha[\bm{\Phi},\bm{\Phi}_q]  =&  -(\Phi^\alpha-\Phi_q^\alpha)\, + \, \frac{1}{2} {}^\alpha\Gamma_{\beta\gamma}[\bm{\Phi}] (\Phi^\gamma-\Phi_q^\gamma)(\Phi^\beta-\Phi_q^\beta) \; +\; \cdots.
 \end{aligned}
\end{align}

The effective action~\eqref{eq:covariant effective action} can be expanded covariantly giving at one- and two-loop levels,
\begin{align}
\Gamma^{(1)}[\bm{\Phi}]=&-\frac{i}{2}\ln\sdet\left( \overrightarrow{\nabla}^{\widehat{\alpha}} S\overleftarrow{\nabla}\!_{\widehat{\beta}}\right)\, ,\label{eq:1 loop eff action}\\[3mm]
\Gamma^{(2)}  [  \bm{\Phi}]   =&   \frac{1}{8} S\overleftarrow{\nabla}_{\!\widehat{\alpha}}\overleftarrow{\nabla}_{\!\widehat{\beta}}\overleftarrow{\nabla}_{\!\widehat{\gamma}}\overleftarrow{\nabla}_{\!\widehat{\delta}}\;{}^{\widehat{\delta}\widehat{\gamma}}\Delta{}^{\widehat{\beta}\widehat{\alpha}}\Delta\nonumber\\
&-(-1)^{\widehat{\gamma}\widehat{\beta}+\widehat{\epsilon}(\widehat{\delta}+\widehat{\beta})}\frac{1}{12}\left(S \overleftarrow{\nabla}_{\!\widehat{\epsilon}}\overleftarrow{\nabla}_{\!\widehat{\gamma}}\overleftarrow{\nabla}_{\!\widehat{\alpha}}\right){}^{\widehat{\alpha}}\Delta^{\widehat{\beta}}\;{}^{\widehat{\gamma}}\Delta^{\widehat{\delta}}\;{}^{\widehat{\epsilon}}\Delta^{\widehat{\zeta}}\left(\overrightarrow{\nabla}_{\!\widehat{\zeta}}\overrightarrow{\nabla}_{\!\widehat{\delta}}\overrightarrow{\nabla}_{\!\widehat{\beta}} S\right).\label{eq:2 loop eff action}
\end{align}
Here, ${}^{\widehat{\alpha}\widehat{\beta}}\!\Delta \equiv
\big( \overrightarrow{\nabla}\!_{\widehat{\alpha}}
\overrightarrow{\nabla}\!_{\widehat{\beta}}\,S\big)^{-1}$ and
${}^{\widehat{\alpha}} \Delta^{\widehat{\beta}} \equiv
\big(\overrightarrow{\nabla}\!_{\widehat{\alpha}}\, S\,
\overleftarrow{\nabla}\!_{\widehat{\beta}}\big)^{-1}$ are rank-2 frame-covariant propagators.
As expected, the one- and two-loop effective actions, \eqref{eq:1 loop eff action} and \eqref{eq:2 loop eff action}, are fully reparametrisation invariant.

The two-loop correction~\eqref{eq:2 loop eff action} to the VDW effective action can also be written as a sum of two covariant Feynman diagrams, i.e.
\begin{equation}
\label{eq:gamma2}
\Gamma^{(2)}  [  \bm{\Phi}]   \; = \;\SolidInfinity+\SolidBasketball \ .
\end{equation}
Note that because of the choice~\eqref{eq:choose C},~$\Gamma^{(2)}   [  \bm{\phi}] ~$ contains only 1PI graphs, whereas other possible one-particle reducible diagrams, such as
\begin{equation}
\SolidDumbbell \,,
\end{equation}
evaluate to zero.

In evaluating the expressions~\eqref{eq:1 loop eff action} and~\eqref{eq:2 loop eff action}, we should use the covariant Feynman rules, which are defined by
\begin{equation}
\SolidNPoint{\alpha_1}{\alpha_2}{\alpha_3}{\alpha_n}=\nabla_{\{\widehat{\alpha}_{1}}\ldots \nabla_{\widehat{\alpha}_n\}}S\,.\label{eq:ferm cov feyn rule}
\end{equation}
Here the notation $\{\cdots\}$ implies supersymmetrisation over the indices, i.e. 
\begin{equation}
\{\alpha_i\cdots\alpha_n\}=\frac{1}{n!}\sum_{P}(-1)^P P[\alpha_i\cdots\alpha_n]\,,
\end{equation}
where $P$ runs over all permutations of the $n$ indices and $(-1)^P$ gives $-1$ when the permutation involves an odd number of fermionic commutations and $+1$ otherwise.

\section{Examples}\label{sec:examples}

\subsection{Single Fermion}

As an explicit example, let us consider a theory with a single scalar field~$\phi$ and a single Dirac fermion field~$\psi$. The most general Lagrangian for such a theory with up to quadratic kinetic terms is
\begin{align}
\begin{aligned}
\L=&\hf k(\phi)\partial_\mu\phi\partial^\mu\phi-\hf h(\phi)\overline{\psi}\gamma^\mu\psi\partial_\mu\phi+\frac{i}{2} g(\phi)\overline{\psi}\gamma^\mu\partial_\mu\psi\\
&-\frac{i}{2}g(\phi)\partial_\mu\overline{\psi}\gamma^\mu\psi-Y(\phi)\overline{\psi}\psi-V(\phi),
\end{aligned}\label{eq:example lagrangian}
\end{align}
where~$k$,~$h$,~$g$,~$Y$ and~$V$ are arbitrary real functions of~$\phi$.

Employing~\eqref{eq:extract k} and~\eqref{eq:extract zeta}, we may derive the kinetic model functions for~\eqref{eq:example lagrangian},
\begin{align}
\begin{aligned}
{}_\alpha k_\beta&=\begin{pmatrix}
k(\phi)&0&0\\
0&0&0\\
0&0&0
\end{pmatrix},\\
\zeta_\alpha^\mu&=\begin{pmatrix}
ih(\phi)\overline{\psi}\gamma^\mu\psi,&g(\phi)\overline{\psi}\gamma^\mu,&g(\phi)\gamma^\mu\psi
\end{pmatrix}.
\end{aligned}
\end{align}
By means of~\eqref{eq:def zeta}, we then obtain
\begin{equation}
\zeta_\alpha=\begin{pmatrix}
ih\overline{\psi}\psi,&g\overline{\psi},&g\psi
\end{pmatrix}
\end{equation}
and~\eqref{eq:define Lambda} yields
\begin{equation}
{}_\alpha \Lambda_\beta=\begin{pmatrix}
k&\hf( g'-ih)\overline{\psi}&\hf(g'+ih)\psi\\
\hf(g'-ih)\overline{\psi}&0&g1_4\\
\hf(g'+ih)\psi&g1_4&0
\end{pmatrix},
\end{equation}
where a prime $'$ indicates differentiation with respect to the field $\phi$.

We may now calculate the field-space metric ${}_\alpha G_\beta$ through the vielbeins as described in Section~\ref{sec:metric}. Solving equation~\eqref{eq:find vielbeins}, we see that the vielbeins for this
theory are 
\begin{equation}
{}_\alpha e^a=\begin{pmatrix}
\sqrt{k}&\frac{g'+ih}{2\sqrt{g}}\psi\,x&\frac{g'-ih}{2\sqrt{g}}\overline{\psi}\,x^{-1}\\
0&\sqrt{g}x&0\\
0&0&\sqrt{g}x^{-1}
\end{pmatrix},\label{eq:example vielbeins}
\end{equation}
where~$x$ is an arbitrary invertible~$4\times 4$ matrix that can depend on both~$\phi$ and~$\overline{\psi}\psi$.\footnote{Note that there exists another solution\begin{equation*}{}_\alpha e^a=\begin{pmatrix}
\sqrt{k}&\frac{g'-ih}{2\sqrt{g}}\overline{\psi}x^{-1}&\frac{g'+ih}{2\sqrt{g}}\psi x\\
0&0&\sqrt{g}x\\
0&\sqrt{g}x^{-1}&0
\end{pmatrix}.\end{equation*} But, as discussed in Section~\ref{sec:metric}, this leads to a metric with the wrong signature.} As discussed in Section~\ref{sec:metric}, the choice of~$x(\phi,\overline{\psi}\psi)$ is irrelevant and will cancel out when we calculate the field-space metric.

Using~\eqref{eq:metric vielbein}, we therefore find the field-space metric to be
\begin{equation}
{}_\alpha G_\beta=\begin{pmatrix}
k-\frac{g^{\prime 2}+h^2}{2g}\overline{\psi}\psi&-\hf(g'-ih)\overline{\psi}&\hf(g'+ih)\psi\\
\hf(g'-ih)\overline{\psi}&0&g1_4\\
-\hf(g'+ih)\psi&-g1_4&0
\end{pmatrix}.\label{eq:example metric}
\end{equation}
The superdeterminant of the metric ${}_\alpha G_\beta$ to be used in the path integral measure is
\begin{equation}
\sdet(G)=\frac{k}{g^8}\;.\label{eq:sdet G}
\end{equation}

Substituting~\eqref{eq:example metric} into~\eqref{eq:connections}, we may calculate the field-space affine connections of the theory, which we find to be
\begin{align}
\begin{aligned}
{}^\phi\Gamma_{\phi\phi}&=\frac{k'}{2k}\;,\\
{}^{\psi_a}\Gamma_{\phi\phi}&=\left[-\frac{h^2+g'^2}{4g^2}+\frac{g''+ih'-\frac{k'}{2k}(g'+ih)}{2g} \right] \psi_a,\\
{}^{\psi_a}\Gamma_{\psi_b\phi}&={}^{\psi_a}\Gamma_{\phi\psi_b}=\frac{g'+ih}{2g}\delta_{ab},\\
{}^{\overline{\psi}_{\dot{a}}}\Gamma_{\phi\phi}&=\left[-\frac{h^2+g'^2}{4g^2}+\frac{g''-ih'-\frac{k'}{2k}(g'-ih)}{2g} \right]\overline{\psi}_{\dot{a}},\\
{}^{\overline{\psi}_{\dot{b}}}\Gamma_{\overline{\psi}_{\dot{a}}\phi}&={}^{\overline{\psi}_{\dot{b}}}\Gamma_{\phi\overline{\psi}_{\dot{a}}}=\frac{g'-ih}{2g}\delta_{\dot{a}\dot{b}},
\end{aligned}\label{eq:example connections}
\end{align}
with all other connections vanishing.

Knowing the field-space connections in~\eqref{eq:example connections},  we may evaluate the field-space Riemann tensor~\cite{dewitt1992supermanifolds}
\begin{align}
\begin{aligned} R^\alpha_{\ \beta\gamma\delta}=&-{}^\alpha\Gamma_{\beta\gamma}\overleftarrow{\partial}_\delta+(-1)^{\gamma\delta}\;{}^\alpha\Gamma_{\beta\delta}\overleftarrow{\partial}_\gamma+(-1)^{\gamma(\beta+\epsilon)}\;{}^\alpha\Gamma_{\epsilon\gamma}\;{}^\epsilon\Gamma_{\beta\delta}-(-1)^{\delta(\epsilon+\beta+\gamma)}\;{}^\alpha\Gamma_{\epsilon\delta}\;{}^\epsilon\Gamma_{\beta\gamma}.
\end{aligned}\label{eq:Riemann} 
\end{align}
In this way, we find that all the components of $R^\alpha_{\ \beta\gamma\delta}$ vanish identically, which implies that the field space described by~\eqref{eq:example metric} is flat. Consequently, the theory~\eqref{eq:example lagrangian} can be made canonical with a suitable field reparametrisation. The reparametrisation in question is
\begin{align}
\begin{alignedat}{2}
\phi&\to\widetilde{\phi}&&=\int_0^\phi\sqrt{k(\phi)}d\phi\,,\\
\psi&\to\widetilde{\psi}&&=\sqrt{g(\phi)}\exp\left(\frac{i}{2}\int_0^\phi \frac{h(\phi)}{g(\phi)}d\phi\right)\psi\,,\\
\overline{\psi}&\to\widetilde{\overline{\psi}}&&=\sqrt{g(\phi)}\exp\left(-\frac{i}{2}\int_0^\phi \frac{h(\phi)}{g(\phi)}d\phi\right)\overline{\psi}\, .
\end{alignedat}\label{eq:canonical transformations}
\end{align}
Introducing the field-space multiplet
\begin{equation}
\widetilde{\Phi}^\alpha=\begin{pmatrix}
\widetilde{\phi},&\widetilde{\psi},&\widetilde{\overline{\psi}}
\end{pmatrix}\,,
\end{equation}
we find
\begin{equation}
\L=\hf\partial_\mu\widetilde{\phi}\partial^\mu\widetilde{\phi}+\frac{i}{2}\widetilde{\overline{\psi}}\gamma^\mu\partial_\mu\widetilde{\psi}-\frac{i}{2}\partial_\mu\widetilde{\overline{\psi}}\gamma^\mu\widetilde{\psi}-\widetilde{Y}(\widetilde{\phi})\widetilde{\overline{\psi}}\widetilde{\psi}-\widetilde{V}(\widetilde{\phi}),\label{eq:canonical tilde lagrangian}
\end{equation}
is canonical as expected. In~\eqref{eq:canonical tilde lagrangian}, we have defined
\begin{align}
\begin{aligned}
\widetilde{Y}(\widetilde{\phi})&=g(\phi) Y(\phi),\\
\widetilde{V}(\widetilde{\phi})&=V(\phi)\,.
\end{aligned}
\end{align}

Notice that, with the choice
\begin{equation}
x=\exp\left(\frac{i}{2}\int_0^\phi \frac{h(\phi)}{g(\phi)}d\phi\right)1_4,
\end{equation}
the vielbeins in~\eqref{eq:example vielbeins} can be identified with the Jacobians of the transformation~\eqref{eq:canonical transformations}. Indeed, we see
\begin{equation}
{}_\alpha e^a=\frac{\overrightarrow{\partial}}{\partial \Phi^\alpha}\widetilde{\Phi}^a.
\end{equation}
As expected, for a flat field space, the vielbeins can be identified with the Jacobian of a transformation and therefore one can move to a field-space frame that is flat everywhere, 
not just locally.

Let us calculate the one-loop effective potential for~\eqref{eq:canonical tilde lagrangian}. Since this Lagrangian is canonically normalised, the field space is trivial and hence we can replace covariant derivatives with partial derivatives. We therefore have
\begin{align}
\begin{aligned}
\overrightarrow{\nabla}\!_{\widehat{\alpha}} S\overleftarrow{\nabla}\!_{\widehat{\beta}}=\begin{pmatrix}
-\square-\widetilde{V}^{\prime\prime}(\widetilde{\phi})-\widetilde{\overline{\psi}}\widetilde{\psi}\widetilde{Y}^{\prime\prime}(\widetilde{\phi})&-\widetilde{\overline{\psi}}\widetilde{Y}^{\prime}(\widetilde{\phi})&\widetilde{\psi}\widetilde{Y}^{\prime}(\widetilde{\phi})\\
\widetilde{\overline{\psi}}\widetilde{Y}^{\prime}(\widetilde{\phi})&0&i\slashed{\partial}-\widetilde{Y}(\widetilde{\phi})\\
-\widetilde{\psi}\widetilde{Y}^{\prime}(\widetilde{\phi})&\big(\!\!-i\slashed{\partial} + \widetilde{Y}(\widetilde{\phi})\big)^{\sf T} &0 
\end{pmatrix}\delta^{(4)}(\bm{x}_\alpha-\bm{x}_\beta).
\end{aligned}\label{eq:example S2}
\end{align}
Plugging this result into~\eqref{eq:1 loop eff action}, we find up to one-loop,
\begin{align}
\begin{aligned}
  \Gamma[\Phi]\ =&\ S[\Phi]-\frac{i}{2}\Tr\ln\bigg\{\square+\widetilde{V}^{\prime\prime}(\widetilde{\phi})
  -\widetilde{\overline{\psi}}\Big[2\big(\widetilde{Y}^{\prime}(\widetilde{\phi})\big)^2
  \big(\!\!-i\slashed{\partial} +\widetilde{Y}(\widetilde{\phi})\big)^{-1} -\,\widetilde{Y}^{\prime\prime}(\widetilde{\phi})\Big]\widetilde{\psi}\,\bigg\}\\
  &+\, i\Tr\ln\Big(\!\!-i\slashed{\partial}
  +\widetilde{Y}(\widetilde{\phi})\Big).
\end{aligned}
\end{align}
This agrees with previous results in the literature (see eq. (8.49) in~\cite{ZinnJustin:1989mi}).

\subsection{Multiple Fermions}

We now generalise the previous example by including~$N$ scalars~$\phi^A$ and~$M$ Dirac fermions~$\psi^X$. The most general Lagrangian derivable from~\eqref{eq:tensor lagrangian} for such a theory with up to quadratic kinetic terms is
\begin{align}
\begin{aligned}
\L=&\hf g^{\mu\nu}k_{AB}(\bm\Phi)\partial_\mu\phi^A \partial_\nu\phi^B-\hf  h_{AXY}(\bm\Phi)\overline{\psi}{}^X\gamma^\mu\psi^Y\partial_\mu\phi^A\\
&+\frac{i}{2}g_{XY}(\bm\Phi)\left(\overline{\psi}{}^X\gamma^\mu\partial_\mu\psi^Y-\partial_\mu\overline{\psi}^X\gamma^\mu\psi^Y\right)\\
&+\frac{i}{2} j_{WXYZ}(\bm\Phi)\overline{\psi}{}^W\!\gamma^\mu\psi^X\left(\overline{\psi}{}^Y\partial_\mu\psi^Z-\partial_\mu\overline{\psi}{}^Y \psi^Z\right)\\
&-Y_{XY}(\bm\Phi)\overline{\psi}{}^X\psi^Y-V(\bm\phi).
\end{aligned}\label{eq:general multi fermi lagrangian}
\end{align}
In the above expression,~$W$,~$X$,~$Y$ and~$Z$ run over the different species of fermion field, so they lie in the interval~${1\leq (W,X,Y,Z)\leq M}$.

The model functions for this theory are
\begin{align}
\begin{aligned}
{}_\alpha k_\beta&=\begin{pmatrix}
k_{AB}&\bm0_{N\times8M}\\
\bm0_{8M\times N}&\bm0_{8M\times 8M}
\end{pmatrix},\\
\zeta_\alpha^\mu&=\begin{pmatrix} ih_{AXY}\,\overline{\psi}{}^X\gamma^\mu\psi^Y\\
g_{YX}\,\overline{\psi}{}^Y\!\gamma^\mu+j_{WYXZ}\overline{\psi}{}^W\!\gamma^\mu\psi^Y\overline{\psi}{}^Z\\
g_{XY}\,\gamma^\mu\psi^Y+j_{WYZX}\overline{\psi}{}^W\!\gamma^\mu\psi^Y\psi^Z
\end{pmatrix},\\
U&=Y_{XY}\overline{\psi}{}^X\psi^Y+V.
\end{aligned}
\end{align}

For illustration,  we consider the case when the Lagrangian~\eqref{eq:general multi fermi lagrangian} contains no terms higher than quadratic order in the fermions. We therefore set $j_{WXYZ}=0$ and assume all other model functions to depend only on the scalar fields. In this case, the Lagrangian takes
on the simpler form
\begin{align}
\begin{aligned}
\L=&\hf g^{\mu\nu}k_{AB}(\bm\phi)\partial_\mu\phi^A \partial_\nu\phi^B-\hf  h_{AXY}(\bm\phi)\overline{\psi}{}^X\gamma^\mu\psi^Y\partial_\mu\phi^A\\
&+\frac{i}{2}g_{XY}(\bm\phi)\left(\overline{\psi}{}^X\gamma^\mu\partial_\mu\psi^Y-\partial_\mu\overline{\psi}{}^X\gamma^\mu\psi^Y\right)\\
&-Y_{XY}(\bm\phi)\overline{\psi}{}^X\psi^Y-V(\bm\phi)\,.
\end{aligned}\label{eq:multi fermi lagrangian}
\end{align}

From~\eqref{eq:def zeta}, we have, for the Lagrangian~\eqref{eq:multi fermi lagrangian},
\begin{equation}
\zeta_\alpha=\left(ih_{AXY}\,\overline{\psi}{}^X\psi^Y,\;\;g_{YX}\,\overline{\psi}{}^Y,\;\;g_{XY}\,\psi^Y\right),
\end{equation}
which, using~\eqref{eq:define Lambda} gives 
\begin{align}
{}_\alpha \Lambda_\beta=&\left(\begin{array}{ccc}
k_{AB}-(h_{AWZ,B}-h_{BWZ,A})\overline{\psi}{}^W\!\!\psi^Z&\hf( g_{ZY,A}-ih_{AZY})\overline{\psi}{}^Z&\hf(g_{YZ,A}+ih_{AYZ})\psi^Z\\
\hf(g_{ZX,B}-ih_{BZX})\overline{\psi}{}^Z&0&g_{YX}1_4\\
\hf(g_{XZ,B}+ih_{BXZ})\psi^Z&g_{XY}1_4&0
\end{array}\right)\label{eq:multi fermi lambda}
\end{align}

We notice that the Lagrangian~\eqref{eq:multi fermi lagrangian} includes a term with a single derivative of a scalar field as did the Lagrangian~\eqref{eq:example lagrangian}. Such a term should be there generically and even appears in free canonically normalised theories if we perform a suitable field redefinition. To make this more explicit, let us consider a field redefinition of the form
\begin{equation}
\psi^X\to\widetilde{\psi}^X=(K(\bm\phi)^{-1})^X_Y\psi^Y.\label{eq:restricted transformation}
\end{equation}
In the new fields, the Lagrangian~\eqref{eq:multi fermi lagrangian} becomes
\begin{align}
\L=&\hf g^{\mu\nu}k_{AB}\partial_\mu\phi^A \partial_\nu\phi^B-\hf  \Big[h_{AXY}K^{*X}_{\,W}K^Y_Z+ig_{XY}\left(K^{*X}_{\,W} \frac{\partial K^Y_Z}{\partial\phi^A}-\frac{\partial K^{*X}_{\,W}}{\partial\phi^A}K^Y_Z\right) \Big]\widetilde{\overline{\psi}}{}^W\gamma^\mu\widetilde{\psi}^Z\partial_\mu\phi^A\nonumber\\
&+\frac{i}{2}g_{XY}K^{*X}_{\,W}K^Y_Z\left(\widetilde{\overline{\psi}}{}^W\gamma^\mu\partial_\mu\widetilde{\psi}^Z-\partial_\mu\widetilde{\overline{\psi}}{}^W\gamma^\mu\widetilde{\psi}^Z\right)-Y_{XY}K^{*X}_{\,W}K^Y_Z\widetilde{\overline{\psi}}{}^W\widetilde{\psi}^Z-V.\label{eq:tilde lagrangian}
\end{align}
We see that, even if originally the theory was canonical with~$h_{AXY}=0$ and~$g_{XY}=\delta_{XY}$, a term with a single derivative of the scalar field will appear after the transformation~\eqref{eq:restricted transformation}.

We may attempt to undo this transformation in order to remove~$h_{AXY}$ with an appropriate field redefinition. If we can find a matrix~$K^X_Y(\bm\phi)$ that satisfies
\begin{equation}
h_{AXY}K^{*X}_{\,W}K^Y_Z=ig_{XY}\left(\frac{\partial K^{*X}_{\,W}}{\partial\phi^A}K^Y_Z-K^{*X}_{\,W} \frac{\partial K^Y_Z}{\partial\phi^A}\right)\,,\label{eq:remove h}
\end{equation}
then this term can be removed by performing the transformation~\eqref{eq:restricted transformation}. However, we see that unlike in the single-fermion case, \eqref{eq:remove h} cannot always be solved in multi-fermion theories, and hence the $h_{AXY}$-dependent term must be considered consistently.

In order to obtain the field-space vielbeins, and hence the metric, for this theory we need to solve~\eqref{eq:find vielbeins} with ${}_\alpha\Lambda_\beta$ given by~\eqref{eq:multi fermi lambda}. This is a very challenging equation to solve in general, and we save such a general solution for future work.

However, we \emph{can} solve this equation for a simple case with only a single bosonic field $\phi$. This allows us to suppress all indices in the bosonic sector. In this case,~\eqref{eq:multi fermi lambda} reduces to
\begin{align}
&{}_\alpha \Lambda_\beta=\left(\begin{array}{ccc}
k&\hf( g'_{ZY}-ih_{ZY})\overline{\psi}{}^Z&\hf(g'_{YZ}+ih_{YZ})\psi^Z\\
\hf(g'_{ZX}-ih_{ZX})\overline{\psi}{}^Z&0&g_{YX}1_4\\
\hf(g'_{XZ}+ih_{XZ})\psi^Z&g_{XY}1_4&0
\end{array}\right).\label{eq:multi fermi single bose lambda}
\end{align}

Solving~\eqref{eq:find vielbeins}, we find the the vielbeins for this theory are
\begin{align}
&{}_\alpha e^a=\begin{pmatrix}
\sqrt{k}&\hf\sqrt{g}^{-1}_{YW}(g'_{ZW}+ih_{ZW})\psi^Z&\hf\sqrt{g}^{-1}_{WY}( g'_{ZW}-ih_{ZW})\overline{\psi}{}^Z\\
0&\sqrt{g}_{YX}&0\\
0&0&\sqrt{g}_{XY}
\end{pmatrix}.
\end{align}
Here $\sqrt{g}_{XY}$ is the matrix square root of $g_{XY}$, satisfying
\begin{equation}
\sqrt{g}_{XZ}\sqrt{g}_{ZY}=g_{XY}
\end{equation}
and $\sqrt{g}^{-1}_{XY}$ is its inverse.

Plugging this into~\eqref{eq:metric vielbein}, we find that the metric for this theory is
\begin{align}
\begin{aligned}
{}_\alpha G_\beta=&\left(\begin{array}{ccc}
k-\hf(g'_{XZ}-ih_{XZ})g^{-1}_{ZW}(g'_{WY}+i h_{WY})\overline{\psi}{}^X\psi^Y&-\hf( g'_{ZY}-ih_{ZY})\overline{\psi}{}^Z&\hf(g'_{YZ}+ih_{YZ})\psi^Z\\
\hf(g'_{ZX}-ih_{ZX})\overline{\psi}{}^Z&0&g_{YX}1_4\\
-\hf(g'_{XZ}+ih_{XZ})\psi^Z&-g_{XY}1_4&0
\end{array}\right)
\end{aligned}\label{eq:multi fermi metric}
\end{align}
Hence, the superdeterminant of the metric, which will appear in the path integral measure, is given by
\begin{equation}
\sdet{(G)}=\frac{k}{(\det g)^8}.
\end{equation}
Notice that~$\sdet(G)=\sdet(\Lambda)$ as expected.

From the metric~\eqref{eq:multi fermi metric}, we could proceed as before and compute, through~\eqref{eq:connections}, the field space connections which enter the VDW effective action. However, such a computation becomes inextricably involved and so we choose to present these results in a future publication.

\section{Discussion}\label{sec:discussion}

We have constructed a field-space supermanifold for theories with fermionic degrees of freedom. We have shown how to equip the field space with a proper metric that can be calculated from the classical Lagrangian. This was achieved through the use of field-space vielbeins. Finally, we have shown that this field-space metric can be used to write down the quantum effective action for fermionic theories in a way that is frame- and reparametrisation-invariant.

The addition of fermionic degrees of freedom makes the identification of the field-space metric more involved. For purely bosonic field theories, the Lagrangian contains terms with two derivatives and the coefficients of such a term readily transform as a rank-2 tensor. However, fermions only appear with one derivative in the Lagrangian and so there is no analogous rank-2 tensor that can be immediately identified with the metric. Thus, whereas for bosonic theories the metric can be calculated directly from the Lagrangian, for fermionic theories the metric must be found indirectly by solving~\eqref{eq:find vielbeins} for the field-space vielbeins.

This difference also means that the relation between the field theory and the geometry of the field space is much more hidden for fermionic theories than it is for bosonic theories. For bosonic theories, the kinetic part of the Lagrangian is proportional to the line element of the field space. Consequently, in the absence of potential terms, the equations of motion of a scalar field theory are the geodesic equations of the field space. 
 In addition, any Noether symmetries of the field theory must obey Killing's equation on the field space~\cite{Finn:2018cfs}.

With the addition of fermions, this connection is no longer evident. The Lagrangian bears no 
direct relation to the line element of the field-space supermanifold and as such, the equations of motion of the fields are not directly related to the geodesic equation for this field space. It would be interesting to examine whether the relation between Noether symmetries of field theory and Killing vectors of the field space still holds in fermionic theories, but such investigations 
lie beyond the scope of the present work.

Nevertheless, the VDW effective action formalism can still be applied and used to define a frame- and reparametrisation-invariant quantum effective action, as stated in~\eqref{eq:covariant effective action}. Hence, it is possible to describe the complete theory with an effective action that is independent of field reparametrisations. This allows us to draw a clear dividing line between the \emph{content} of a theory (i.e. the physical observations it predicts) and its \emph{representation} (how we choose to write the theory down).

We have so far considered only scalar-fermion theories with up to quadratic kinetic terms. An interesting generalization would be to consider theories with higher derivatives. If the techniques developed in this paper were to be applied directly to such theories they would lead to a Finslerian metric~\cite{Finsler}, i.e. one that depends on both the fields and their derivatives. Investigating the effects of such a metric would be a worthwhile task, but is beyond the scope of this current paper.

Other possible directions for future work include the study of field-space torsion or non-metricity. We have so far taken the field-space connections to be the Christoffel symbols for the metric and so have excluded such possibilities. However, it may be advantageous to define torsionful or non-metric field spaces for certain theories. We note that this would only be possible for theories with a curved field space. For theories with a flat field space, it is always possible to define a parametrisation for which the kinetic terms are canonical, at which point all field-space effects should disappear.

The VDW effective action has been applied to scalar theories~\cite{Vilkovisky:1984st,Vilkovisky:1984un,Gell-Mann1960,DeWitt:1985sg,Rebhan:1986wp,Ellicott:1987ir,GrootNibbelink:2000vx,GrootNibbelink:2001qt,vanTent:2003mn}, gauge theories~\cite{Vilkovisky:1984st,Vilkovisky:1984un,Kunstatter:1986qa,DeWitt:1980jv,DeWitt:1988dq,Fradkin:1983nw} and gravity~\cite{Vilkovisky:1984st,Vilkovisky:1984un,Finn:2019aip,DeWitt:1967ub,DeWitt:1967uc,DeWitt:1967yk,Huggins:1986ht,Lavrov:1988is,Buchbinder:1988np,Steinwachs:2013tr,Moss:2014nya,Bounakis:2017fkv}. The addition of fermions in this paper completes the geometrisation of QFTs for a wide range of theories. We are now in a position to express any theory that includes scalar fields, gravity, gauge bosons, or fermions with up to quadratic kinetic terms in a frame- and repara\-metrisation-invariant manner. It should therefore be straightforward to construct a field-space supermanifold, not only for the generic models presented in this paper, but for realistic theories of high energy physics, including the Standard Model.

\begin{acknowledgements} The authors would like to thank Sam Brady, Owen Goodwin and Matthew Kellett for useful comments and discussion on the mathematical properties of supermanifolds. KF is supported by the University of Manchester through the President's Doctoral Scholar Award. The work of SK was supported by the ERC grant 669668 NEO-NAT. The work of AP is supported by the Lancaster -- Manchester -- Sheffield Consortium for Fundamental Physics under STFC research grant ST/L000520/1.  \end{acknowledgements}

\begin{appendices}
\renewcommand*{\thesubsection}{\alph{subsection}}
\renewcommand*{\thesubsubsection}{(\roman{subsubsection})}

\section{Failed Attempt: The Covariant Metric}
\label{sec:covariant metric}
In this appendix we consider an alternative definition of the metric
\begin{equation}
{}_\alpha\widetilde{G}_\beta={}_\alpha k_\beta+\overrightarrow{\nabla}_{\!\alpha}\zeta_\beta+(-1)^{\alpha+\beta+\alpha\beta}\overrightarrow{\nabla}_{\!\beta}\zeta_\alpha,\label{eq:covariant metric}
\end{equation}
which is a supersymmetric rank-2 field-space tensor. We show explicitly why such a definition does not work.

Note that connection terms on the RHS of~\eqref{eq:covariant metric} do not cancel as they do in~\eqref{eq:define lambda} and thus the metric~${}_\alpha\widetilde{G}_\beta$ appears on both sides of this implicit equation. Such an equation is difficult to solve in general, but we can still solve it in certain cases.

We consider solving~\eqref{eq:covariant metric} for the example theory of a single scalar field $\phi$ and a single Dirac fermion $\psi$ shown in~\eqref{eq:example lagrangian}. We take as an ansatz
\begin{align}
&{}_\alpha \widetilde{G}_{\beta}=\begin{pmatrix}
H(\phi)+A(\phi)\overline{\psi}\psi&B(\phi)\overline{\psi}+C(\phi)\psi&D(\phi)\overline{\psi}+E(\phi)\psi\\
-B(\phi)\overline{\psi}-C(\phi)\psi&0&G(\phi)\\
-D(\phi)\overline{\psi}-E(\phi)\psi&-G(\phi)&0
\end{pmatrix}.\label{eq:metric ansatz}
\end{align}
This is the most general ansatz compatible with the fermionic structure of the metric. After plugging~\eqref{eq:metric ansatz} into~\eqref{eq:covariant metric} and performing some algebra, we find that the solution is
\begin{equation}
{}_\alpha \widetilde{G}_\beta=\begin{pmatrix}
k(\phi)+A(\phi)\overline{\psi}\psi&B(\phi)\overline{\psi}&-B(\phi)\psi\\
-B(\phi)\overline{\psi}&0&0\\
B(\phi)\psi&0&0
\end{pmatrix}\,,\label{eq:covariant example metric}
\end{equation}
where~$A(\phi)$ and~$B(\phi)$ are arbitrary functions of~$\phi$.

We can immediately see several problems.
\begin{enumerate}
\item The metric~\eqref{eq:covariant example metric} contains arbitrary functions~$A(\phi)$ and~$B(\phi)$  and as such, it is not uniquely defined by~\eqref{eq:covariant metric} or the Lagrangian~\eqref{eq:example lagrangian}.
\item The metric~\eqref{eq:covariant example metric} does not reduce to the flat metric in the canonical case with~$h=0$, $g=1$ and~$k=1$.
\item The metric~\eqref{eq:covariant example metric} is singular with~$\sdet(\widetilde{G})=\infty$.
\item The metric~\eqref{eq:covariant example metric} has no dependence on the model functions~$h$ and~$g$, and so it is disconnected with the fermionic part of the theory 
that it should describe.
\end{enumerate}

We can find one more problem with the metric~\eqref{eq:covariant metric} if we consider adding to the Lagrangian a total derivative term
\begin{equation}
\L\to\widetilde{\L}=\L+\frac{i}{2}\partial_\mu\left(t(\phi)\overline{\psi}\gamma^\mu\psi\right).\label{eq:add t}
\end{equation}
Such a boundary term will drop out of the classical action when integrated and thus will not affect the results of the theory. However, if we repeat the above derivation for~$\widetilde{\L}$, we find the metric defined by~\eqref{eq:covariant metric} is
\begin{align}
\begin{aligned}
{}_\alpha \widetilde{G}_\beta=\begin{pmatrix}
k+A\overline{\psi}\psi&B\overline{\psi}&\left[-B+2\frac{(B-ih-t')t}{g+t}\right]\psi\\
-B\overline{\psi}&0&-2t\\
\left[B-2\frac{(B-ih-t')t}{g+t}\right]\psi&2t&0
\end{pmatrix},
\end{aligned}\label{eq:general AB metric}
\end{align}
where~$A$ and~$B$ are again arbitrary functions of~$\phi$. We observe that~\eqref{eq:general AB metric} depends strongly on the function~$t(\phi)$ even though, as we argued above, this function is irrelevant to the physics of the theory. Note that, in contrast, the tensor~${}_\alpha\Lambda_\beta$ defined in~\eqref{eq:define Lambda} is not affected by the transformation~\eqref{eq:add t} and therefore the metric defined in Section~\ref{sec:metric} does not suffer from this issue.

We conclude from this exercise that~\eqref{eq:covariant metric} does not constitute a proper definition of the field-space metric for fermionic theories.


\end{appendices}

\end{document}